# Stopping microfluidic flow

Mehmet Akif Sahin[1], Muhammad Shehzad[1], Ghulam Destgeer*[1]

[1] Department of Electrical Engineering, School of Computation, Information and Technology (CIT), Control and Manipulation of Microscale Living Objects, Central Institute for Translational Cancer Research (TranslaTUM), Technical University of Munich, Munich, Germany, 81675
*corresponding author: Ghulam Destgeer, PhD; Einsteinstraße 25, 81675, Munich; +49.89.4140.9037; ghulam.destgeer@tum.de

## Significance statement:

Inside a microfluidic circuit, a significantly high flow resistance and pressure gradient across the circuit length results in a non-uniform expansion of the microchannel walls and connecting tubings. Upon removal of the pressure gradient, the microfluidic circuit should relax back to a state with uniform expansion across its length, which results in a residual flow with $O$(mm/s) velocity inside the microchannel that decays over a time scale of seconds. Three different stop-flow configurations were benchmarked to effectively diminish this residual velocity as quickly as possible by readily neutralizing the pressure gradient and homogenizing the circuit compliance. The ability to rapidly stop the microfluidic flow will be transformative for the fields of additive manufacturing, flow lithography, and 3D printing.

## Abstract:

We present a cross-comparison of three stop-flow configurations—such as low-pressure (LSF), high-pressure open-circuit (OC-HSF), and high-pressure short-circuit (SC-HSF) stop-flow—to rapidly bring a high flow velocity within a microchannel to a standstill. The average velocities inside the microchannels were reduced from >1 m/s to <10 µm/s within 2s of initiating the stop-flow. The performance of the three stop-flow configurations was assessed by measuring the residual flow velocities within microchannels having three orders-of-magnitude different flow resistances. The LSF configuration outperformed the OC-HSF and SC-HSF configurations within the high flow resistance microchannel, and resulted in a residual velocity of <10 µm/s. The OC-HSF configuration resulted in a residual velocity of <150 µm/s within a low flow resistance microchannel. The SC-HSF configuration resulted in a residual velocity of <200 µm/s across the three orders-of-magnitude different flow resistance microchannels, and <100 µm/s for the low flow resistance channel. We hypothesized that the residual velocity resulted from the compliance in the fluidic circuit, which was further investigated by varying the elasticity of the microchannel walls and the connecting tubing. A numerical model was developed to estimate the expanded volumes of the compliant microchannel and connecting tubings under a pressure gradient and to calculate the distance traveled by the sample fluid. A comparison of the numerically and experimentally obtained traveling distances confirmed our hypothesis that the residual velocities were an outcome of the compliance in the fluidic circuit. Therefore, a configuration with minimal fluidic circuit compliance resulted in the least residual velocity.

**Graphical Abstract:**

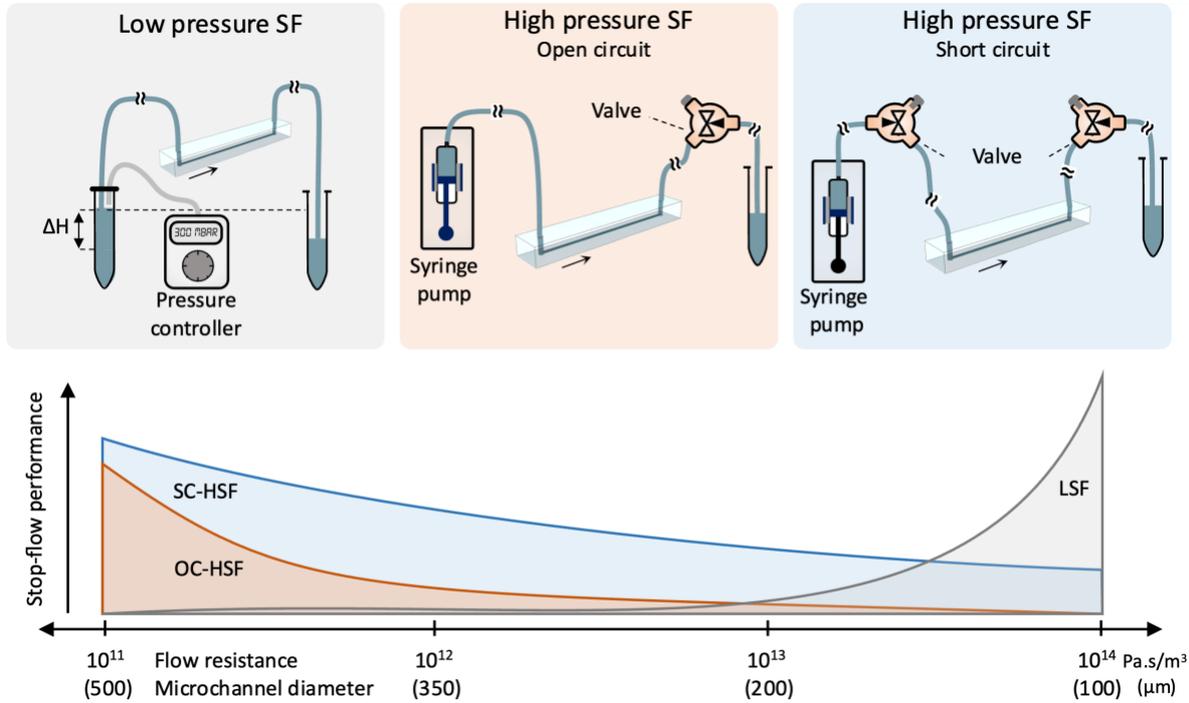

**Introduction:**

Controlling fluid flow inside a microfluidic channel with the possibility of an on-demand spatiotemporal modulation has important implications for 3D printing,[1]–[3] flow lithography,[4]–[7] rapid pulsating flow generation,[8]–[10] tunable concentration gradients formation,[11]–[13] etc. For example, tunable switching of an extruding media from a 3D printing nozzle is critical to building discrete and complex structures, where the ability to *control, i.e., on-demand start and stop,* a certain flow stream defines the efficacy of the printer head. Here, the deterministic nature of the laminar flow within microfluidic channels offers a controllable regimen for a desired printing application, where a viscoelastic printable material helps to achieve faster flow switching response time.[2], [3] Similarly, a rapidly pulsating flow of partially miscible aqueous two phases with minimal interfacial tension can be used to generate water-in-water droplets, which would not have been possible using the conventional droplet generation strategies in the absence of noticeable Rayleigh plateau instability.[8] Lastly, cyclic flow control of structured fluidic streams in a 'stop flow lithography' process is an essential prerequisite to manufacturing multi-material, anisotropic, 3D microparticles with numerous applications across fields.[14]–[18]

Controlling a low Reynolds number ($Re \ll 1$) laminar Stokes flow with insignificant inertial effect and minimal pressure gradient is facile compared to a high Reynolds number ($1 < Re < 500$) laminar inertial flow with relatively high velocity, pressure gradient, and momentum, which requires extra measures to precisely modulate the flow. The inescapable inertial effects such as Dean vortices and wall- and shear-induced lift forces can be used to your advantage to sculpt the flow or manipulate suspended micro-objects at very high throughputs[19], [20] without breaking the deterministic nature of laminar flow regime.[21] Similarly, a high viscosity flow of multiple streams with low diffusivity and high Péclet number ($Pe$) can limit the diffusive mixing of streams for longer lengths of the microchannels.[22]–[24] Despite several advantages, inertial flows with high $Re$ and $Pe$ require a very high flow rate and steep pressure gradient that lead to high compliance in the microfluidic channel and long momentum diffusion time, which makes it challenging to *control* the flow in a pulsatile manner within reasonable time periods.[18] A deep understanding of the limits of *inertial flow control* is critical to developing new strategies for effectively applying microfluidics in emerging fields of additive manufacturing, flow lithography, 3D printing, etc. In this work, we have investigated three different flow-control configurations to find the favorable conditions for stopping the high viscosity flows inside the microfluidic channels within the shortest possible times by rapidly neutralizing pressure gradients of up to 2.5 bar (**Fig. 1**).

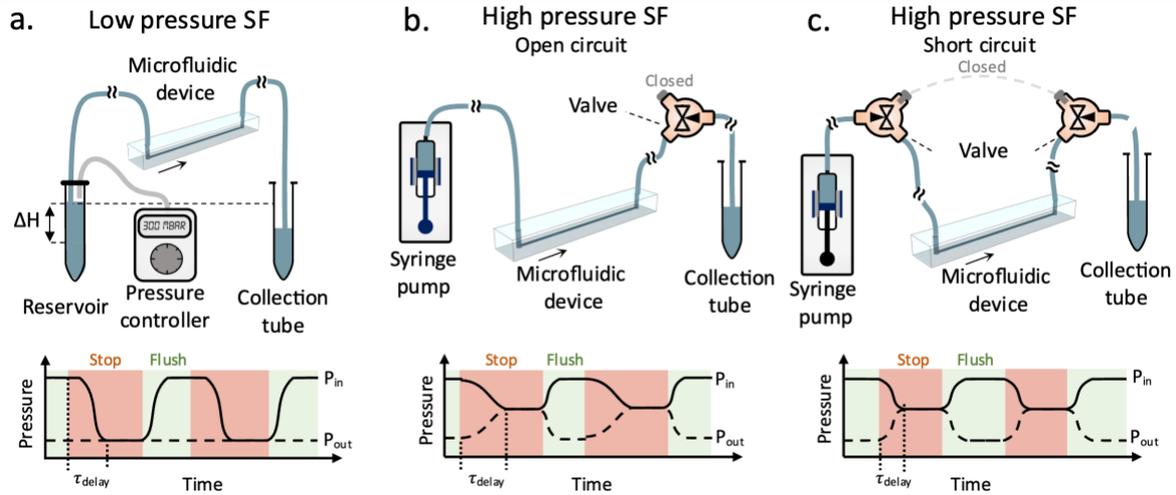

**Figure 1. Three stop-flow configurations.** (a) Low-pressure stop flow (LSF) utilizes a compressed air source to rapidly actuate flow within a liquid reservoir to start or stop the flow within the microchannel, where the flow only stops as the $P_{in} \rightarrow P_{out}$ after a certain delay $\tau_{delay}$. (b-c) In high-pressure stop flow (HSF), a syringe pump pushes the liquid sample through a microchannel. (b) A single valve at the outlet in the open circuit HSF (OC-HSF) or (c) two valves at the inlet and outlet of the microchannel in the short circuit HSF (SC-HSF) enable the flow stop when operated in sync with the syringe pump, where $P_{in}$ and $P_{out}$ meet at an intermediate value as the flow is stopped.

A fluid flow, which emerges as a consequence of the pressure gradient present across a microchannel length, stops as the pressure gradient is neutralized by bringing the inlet and outlet pressures to atmospheric pressure (i.e., low-pressure stop flow) or by matching the inlet and outlet pressures at a higher level above atmospheric pressure (i.e., high-pressure stop-flow). In a low-pressure stop flow (LSF) configuration, a compressed air source connected with a pressure regulator pushes a liquid sample in a reservoir connected to the inlet and generates a flow through the microchannel to the collection tube maintained at atmospheric pressure (Fig. 1a).[15] The pressure regulator rapidly alters the air pressure at the microchannel inlet to start or stop the flow as the air experiences minimal flow resistance moving in or out of the reservoir. High-pressure stop flow (HSF) configurations use a syringe pump to push the liquid sample through a microchannel instead of a compressed air source (Fig. 1b-c). In an open-circuit HSF (OC-HSF), the flow is stopped by simultaneously turning the syringe pump off and closing a valve downstream of the microchannel outlet, which homogenizes the pressure throughout the fluidic circuit (Fig. 1b).[18], [22], [23] In a short-circuit HSF (SC-HSF), an additional valve upstream of the microchannel is used to disengage the syringe pump from the fluidic circuitry, where the flow is stopped by simultaneously closing the two valves and stopping the syringe pump (Fig. 1c). This prevents any perturbations from the syringe pump to be re-laid to the microchannel. In an ideal scenario, the flow should stop immediately as the pressure source is removed from the fluidic circuit; however, due to the compliance of the microchannel and connecting tubings, also known as fluidic capacitance, it takes a certain delay time ($\tau_{delay}$) before a residual flow velocity ($v_r$) within the microchannel approaches zero. This $v_r$ is used to benchmark the three stop-flow configurations described above.

In this study, we have investigated three microchannels (hydraulic diameter $D_h$: 100, 200, and 500 µm with square cross sections) with two different materials (glass and polydimethylsiloxane (PDMS)) offering a range of flow resistances ($10^{11}$-$10^{14}$ Pa.s/m$^3$) and fluidic capacitances to compare the three stop-flow configurations (LSF, OC-HSF, and SC-HSF). A range of steady-state flow rates (0.1-200 µl/s) was realized by varying the pressure drops (0.05-2.5 bar) across microchannels with different hydraulic diameters. The average residual flow velocities ($v_r$) within microchannels were measured by tracing fluorescent microparticles suspended in the media by using an in-house built algorithm detecting individual particles' positions and velocities during the delay time ($\tau_{delay}$). Three different materials (polyvinyl chloride (PVC), polytetrafluoroethylene (PTFE), and steel) of the connecting tubing were used to understand their influence on the compliance of the fluidic circuit. The role of sample viscosity (1-24 mPa.s) contributing to the hydraulic flow resistance and flow rates affecting the pressure drop across the channel length was also studied. Lastly, a numerical model was developed to explain the possible reasons behind a residual flow velocity and understand the contributions of individual elements of the fluidic circuit that lead to a delay time prior to the flow stoppage.

**Results and discussion:**

**Comparison of stop-flow configurations:**

We compare the three stop-flow configurations by measuring the residual velocities ($v_r$) within the microchannels as the flow is actively stopped by removing the pressure gradient ($\Delta P$) across the fluidic circuit (**Fig. 2**). To limit the compliance of the fluidic circuit, we used a glass syringe (for HSF configurations only), and glass microchannels of different cross sections with PTFE tubings connected at the inlet and outlet ports by blunt end needles (**Table S1-S2**). We measured the $v_r$ by tracing 10 µm diameter fluorescent microparticles for three stop-flow cycles to take an average velocity value over a specific time interval (**Fig. S1**).

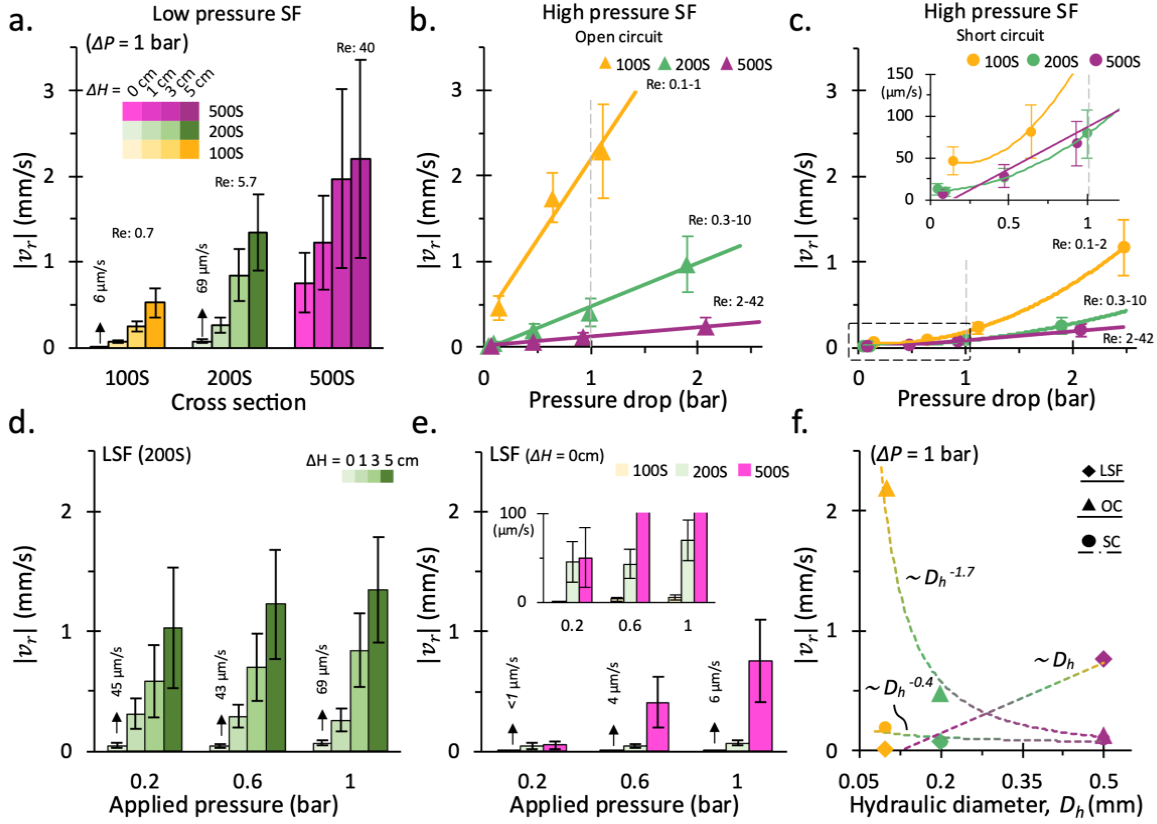

**Figure 2. Comparison of low-pressure (LSF), high-pressure short circuit (SC-HSF), and high-pressure open circuit (OC-HSF) stop flow configurations.** (a) Residual velocity ($v_r$) in the LSF method increases with the height difference ($\Delta H$) from 0 cm to 5 cm, and the microchannel diameter from 100S to 500S for a given inlet pressure of 1 bar. In the OC-HSF (b) and SC-HSF (c), $v_r$ increases with the starting pressure gradient ($\Delta P$) but shows a decreasing trend with the increasing microchannel diameter. Lines represent linear fits to OC-HSF data (b) and polynomial fits (second order) to SC-HSF data (c). (d) The LSF method is independent of the applied inlet pressure for the 200S microchannel. (e) The $v_r$ in the 500S microchannel is significantly higher than that in the 100S and 200S microchannels at the higher inlet pressures (0.6 and 1 bar; and $\Delta H$ = 0 cm) in the LSF method as the contribution of a rapidly evolving $\Delta H$ becomes significant for 500S microchannel. For inlet pressure of 0.2 bar, the $v_r$ value for 500S microchannel is also dropped significantly due to slowly changing $\Delta H$ over multiple cycles. (a-e) Data points represent means of multiple cycles and particle replicates ($N_{cycle}$ = 3, $N_{particle}$ > 100), and error bars show standard deviations. (f) For a given $\Delta P$ (1 bar), the $v_r$ increases linearly with the microchannel cross-sectional area in the LSF method, The SC-HSF method results in relatively much lower values of $v_r$ compared to the OC-HSF and LSF methods.

**LSF is sensitive to the hydrostatic pressure fluctuations and performs well with smaller diameter microchannel:**

We observed that the $v_r$ in the LSF configuration is strongly influenced by the hydrostatic pressure fluctuations within the microchannel arising due to the height difference ($\Delta H$) between the sample's free surfaces at the inlet reservoir tube and the outlet collection tube (Fig. 1a and Fig. 2a). As the inlet and outlet free surface levels are meticulously matched, i.e. $\Delta H \cong 0$ cm, to remove any hydrostatic pressure prior to the start of the stop-flow cycles, the average $v_r$ of 6 µm/s is measured within the 100S microchannel during the stop-phase of the cycle corresponding to an inlet pressure of 1 bar during the flush-phase. The $v_r$ increases to 69 µm/s and ~750 µm/s for the 200S and 500S microchannels, respectively. A significantly higher $v_r$ value for the 500S microchannel can be attributed to 625x higher flow rate ($Q = \Delta P/R_h$; flow resistance $R_h \propto D_h^{-4}$), compared to the 100S channel, attained during the flush-phase for the given inlet pressure of 1bar. This leads to 625x higher free surface level in the collection tube for the 500S channel. Therefore, a two orders of magnitude higher $v_r$ is observed due to the resultant hydrostatic pressure and less flow resistance for the larger 500S channel, even though the compressed air inlet pressure source is already disengaged from the fluidic circuit. As the $\Delta H$ is intentionally set to be >0 cm prior to the stop-flow cycles, we observed an increasing trend of $v_r$ against $\Delta H$ for all the microchannels. For example, $v_r$ increases from 6 µm/s to ~0.5 mm/s for the 100S channel as $\Delta H$ is increased from 0 cm to 5 cm. A similar trend continues for the 200S and 500S channels with relatively higher $v_r$ values. For the larger channels, a much higher standard deviation is recorded (e.g. $CV$ >50% for 500S channel with $\Delta H \cong 5$ cm) because of the very different residual velocities measured over the three stop-flows cycles as a continuously evolving hydrostatic pressure results in a variable $v_r$ in each cycle. Particularly, for the 500S channel with significantly higher flow rates, most of the sample volume (~10 ml) in the reservoir tube is consumed within the three stop-flow cycles, thereby altering $\Delta H$ by ~10 cm and resulting in a very different hydrostatic pressure every cycle.

Due to the open atmospheric pressure boundary conditions at the inlet and outlet of the microchannels, the excessive capacitive volume built within the fluidic circuit can easily drain away at the ports of the high flow resistance 100S microchannel without significantly disturbing the flow within the microchannel. However, the $v_r$ within 200S and 500S microchannels stays high as the relatively lower flow resistances, i.e., 16x and 625x lower than the 100S microchannel, respectively, provide less viscous damping to the residual flow produced due to the hydrostatic pressure (Fig. 2a). For example, $v_{r,200S} \approx$ ~4 x $v_{r,100S}$ and $v_{r,500S} \approx$ ~20 x $v_{r,100S}$ when $\Delta H \cong 1$ cm and the inlet pressure is 1 bar. The high $v_r$ values, particularly for larger-diameter microchannels, highlight the vulnerability of the LSF configuration to hydrostatic pressure fluctuations. An increasing trend in the $v_r - \Delta H$ barplot observed for different microchannels at 1 bar (Fig. 2a) is also preserved for different applied pressures of 0.2 bar and 0.6 bar (Fig. 2d). At higher inlet pressures (>1 bar), the reservoir height would shift so dramatically due to the high flow rates through the microchannel that it would be challenging to present the results, therefore, the inlet pressure is limited to 1 bar in these experiments. The $v_r$ is not significantly affected by the applied inlet pressure (0.2-1 bar) at any given $ΔH$ value (0-5 cm) for the LSF configuration tested with 200S microchannel, which is consistent with the earlier report (Fig. 2d).[15] However, a comparison of $v_r$ values between different microchannels for variable inlet pressures (0.2-1 bar) and $ΔH ≅$ 0cm highlights that the $v_r$ values in 100S and 200S channels are relatively stable but in the 500S channel $v_r$ increases significantly from ~0.05 mm/s at 0.2 bar to ~0.75 mm/s at 1 bar (Fig. 2e). This contradicts the observation made with the smaller channels and indicates a pressure-dependent stop-flow behavior of the larger diameter 500S channel. A non-zero $v_r$ when $ΔH ≅$ 0 cm is attributed to fluidic capacitance of the circuit but more importantly an increasing $v_r$ with inlet pressure for 500S channel is a result of an ever-evolving reservoir free-surface height as the liquid sample is consumed during the experiment (Fig. 2e). For perspective, we can theoretically estimate that a height difference $ΔH ≅ ±1$ mm (μ = 10 mPa.s, ρ = 1060 kg/m$^3$) can result in $v_r ≅$ 7 μm/s, 30 μm/s, and 80 μm/s for the 100S, 200S, and 500S microchannels, respectively, or a minor pressure fluctuation $ΔP ≅$ 1 mbar within the channels can generate relatively 10x higher $v_r$ values compared to $ΔH ≅ ±1$ mm. Since the reservoir height shift $ΔH$ is $O(cm)$ for the 500S channel with much lower fluidic resistance to damp the residual flow, its contribution to the $v_r$ cannot be ignored. The LSF configuration stops the flow better (or results in lower $v_r$) for microchannels with high flow resistances and smaller hydraulic diameters but this method is very sensitive to hydrostatic perturbations for larger diameter channels.

**SC-HSF disengages the syringe pump to realize better stop flow performance compared to OC-HSF:**

In the HSF configurations, the $v_r$ increases with the $ΔP$ for all the microchannel diameters (Fig. 2b and 2c). Here, for a better comparison between stop-flow configurations tested with microchannels of variable flow resistances, we first correlated the syringe pump flow rates ($Q$) with the inlet pressures (same as $ΔP$), and used the latter to present the results associated with the HSF configurations (**Fig. S2**a). Using the OC-HSF stop flow configuration for a 100S microchannel, the average $v_r$ increases from ~0.47 mm/s to ~2.23 mm/s as the $ΔP$ increase from ~0.14 bar to ~1.1 bar (Fig. 2b). In a 500S microchannel, using the same method, the maximum $v_r$ is measured as ~0.24 mm/s at $\Delta P$ of ~2.1 bar, which clearly demonstrates better stop-flow conditions within the larger microchannel even at a much higher $\Delta P$. The residual velocities are further reduced by implementing the SC-HSF configuration, where the maximum $v_r$ value of ~1.19 mm/s is recorded for the 100S microchannel at $\Delta P$ of ~2.5 bar (Fig. 2c). The $v_r$ values achieved using the SC-HSF method are approximately an order of magnitude lower than those measured in the OC-HSF method, e.g., ~0.23 mm/s vs ~2.23 mm/s at $\Delta P$ ~1.1 bar, respectively. For $\Delta P$ < 1 bar, the $v_r$ < 0.2 mm/s is recorded for all the microchannels (Fig. 2c, inset). A linear increment of $v_r$ with $\Delta P$ in the OC-HSF can be attributed to the compliance in the syringe holder ($\Delta L$), where a displaced syringe volume ($\Delta V$) due to the rubber padding deformation and relaxation is linearly correlated with the force applied to the plunger (i.e., $\Delta V \sim \Delta L$) (Fig. 2b and **Fig. S3**). The effect of radial compliance of the glass syringe (elastic modulus $E$ ~ 70 GPa) would be minimal, therefore, ignored. At high inlet pressure, the connecting tubing also expands, however, the compliance of the syringe holder overshadows the compliance of the inlet tubing. A quadratic growth of $v_r$ noted in the SC-HSF, when the syringe pump is disengaged from the fluidic circuit, can be attributed to the areal relationship between the increasing $\Delta P$ and expanded cross-section area ($\Delta D^2$) of the connecting tubing ($E$ ~ 0.4 GPa) (Fig. 2c). The expanded inlet tubing adds a volume in the fluidic circuit (i.e., $\Delta V \sim \Delta P \sim \Delta D^2$), which, when dispended through the microchannel upon the removal of $\Delta P$, results in the quadratic $v_r - \Delta P$ curves.

For a given $\Delta P$ of 1 bar, the $v_r$ shows a decreasing trend against the microchannel cross-sectional area ($S$) for the OC-HSF and SC-HSF configurations (Fig. 2f). For the LSF configuration, the $v_r$ increases with $S$ whereas the lowest $v_r$ is recorded for the 100S channel among all the configurations. The $v_r$ values are comparable between the LSF and SC-HSF methods for the 200S microchannel. However, the SC-HSF configuration performs better than the LSF method for the larger 500S microchannel. The OC-HSF results in the highest $v_r$ values for 100S microchannel compared to other configurations, but performs similarly to the SC-HSF configuration for the larger 500S channel.

**High viscosity flow and stiffer tubing material result in lower residual velocities:**
The residual velocity $v_r$ measured within a 200S glass microchannel using SC-HSF for variable sample viscosities showed an increasing trend against $\Delta P$, where high viscosity (~24 mPa.s) pure polymer flow resulted in an order of magnitude lower $v_r$ values compared to low viscosity (~1 mPa.s) water flow for any given value of $\Delta P$ (**Fig. 3a,** Fig. S2b, and Table S3). A 1:1 mixture of the pure polymer solution in water (50% polymer) led to an intermediate sample viscosity (~10 mPa.s), which followed a $v_r - \Delta P$ curve between the two extreme cases. Since the viscosities of the three samples were quite different from each other, each sample was initially pumped at a very different flow rate during the flush-phase of the stop-flow cycle to reach similar $\Delta P$ values following the Hagen-Paussillie equation ($\Delta P = QR_h$) that resulted in a wide range of Reynolds number ($Re = \rho UD/\mu$). For example, the low viscosity (1 mPa.s) sample reached a $Re \cong 500$ ($v_{flush} \cong 2.5$ m/s) as the inlet pressure was set at 1 bar, whereas, the $Re$ for the high viscosity (~24 mPa.s) sample barely crossed 1 ($v_{flush} \cong 0.13$ m/s) even when the inlet pressure was above 2 bar (Fig. 3a). These two orders of magnitude lower $Re$ values for high viscosity sample during the flush-phase resulted in significantly lower $v_r$ values (≤0.1 mm/s) during the stop-flow, which can be attributed to the high viscous-damping of accumulated fluidic capacitance due to compliance of the tubing and channel walls. In contrast, the residual velocity for low-viscosity water flowing at a high flow rate ($Re \cong 500$) stayed above 2 mm/s in the absence of high viscous damping. The 2nd-order polynomial fittings to the $v_r - \Delta P$ curves confirmed that the residual velocity $v_r$ mainly resulted from the tubing compliance. It is interesting to note that at a constant $\Delta P$, the $v_r$ decreases with sample viscosity $\mu$, following an inverse power law decay: $v_r \sim \mu^{-1.29}$ (**Fig. S4**).

We pumped the 50% polymer sample (~10 mPa.s) through a 200S glass microchannel using the SC-HSF configuration to plot the $v_r - \Delta P$ curves except that the inlet tubings of variable elastic modulus were used, i.e., PVC (~10 MPa), PTFE (~0.4 GPa), and steel (~180 GPa) (Fig. 3b and Table S2). The $v_r - \Delta P$ curve for the PTFE tubing follows a 2nd order polynomial fit due to the radial compliance of the tubing. Stiffer steel tubing accumulated significantly less fluidic capacitance compared to the PTFE tubing, which resulted in a relatively lower slope of a linear $v_r - \Delta P$ curve. In contrast, a comparatively softer PVC tubing with respect to PTFE resulted in a higher slope of the 2nd order $v_r - \Delta P$ curve, highlighting higher compliance of the PVC tubing in the radial direction. The inlet tubing was cut short to a minimum length, and the inlet valve was placed next to the inlet port of the microchannel, to realize an experiment without the tubing effect. The resulting $v_r - \Delta P$ curve for the without-tubing case was very similar to that obtained using steel tubing with linear fit, which indicates a small (yet non-zero) slop of the $v_r - \Delta P$ curve. The non-zero $v_r$ values for the steel tubing and no-tubing cases originate from the compliance of the blunt-end connecting needles or valves, and are not due to the tubing compliance.

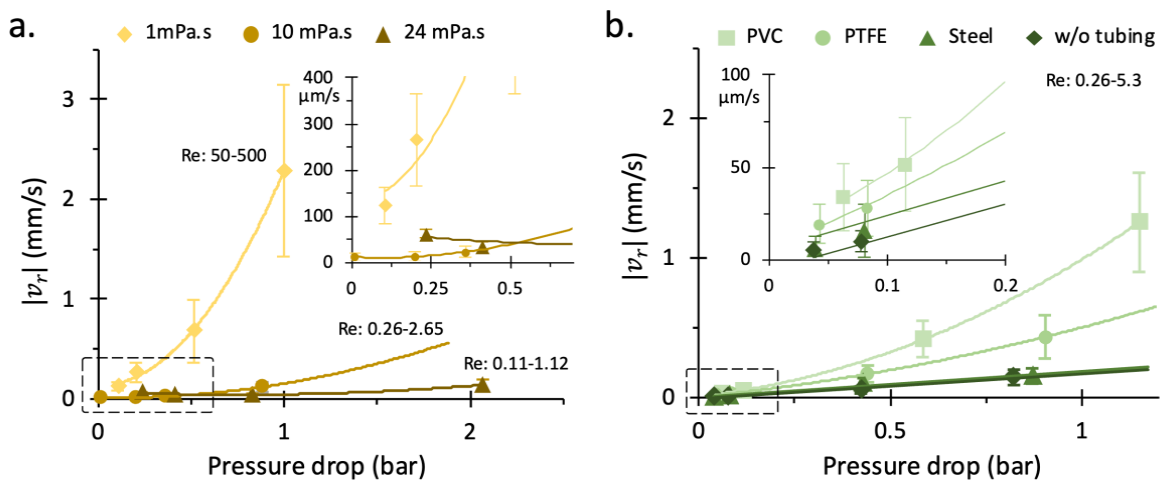

**Figure 3. Effect of sample viscosity and inlet tubing material to the stop flow residual velocity within a 200S glass microchannel using the SC-HSF configuration.** (a) The rise in $v_r$ with $\Delta P$ is most prominent for a low viscosity (1 mPa.s) sample, where a high-velocity flow during the flush phase of the stop-flow cycle can lead to $Re$ of 50-500. For an order of magnitude higher viscosities (10 and 24 mPa.s), two orders of magnitude lower starting $Re$ resulted in significantly lower $v_r$ values. (b) As the elastic modulus of inlet tubing is increased, i.e., from PVC (~10 MPa) and PTFE (~0.4 GPa) to steel (~180 GPa), the slope of $v_r - \Delta P$ curves decreased. The $v_r - \Delta P$ curve for the without-tubing case is very similar to the steel tubing due to the minimal compliance of steel. Data points represent means of multiple cycles and particle replicates ($N_{cycle}$ = 3, $N_{particle}$ > 100), and error bars show standard deviations.

**Effect of microchannel material on stop-flow performance:**

The residual velocity $v_r$ is dominated by the compliance of the syringe holder for the OC-HSF configuration, therefore, the contribution of a PDMS microchannel to the fluidic capacitance and the resultant $v_r$ value is overshadowed (**Fig. 4a**). The $v_r - \Delta P$ curves for the softer PDMS microchannel ($E$ = ~0.75 MPa) are not significantly different from those for the stiffer glass microchannel ($E$ = 70 GPa). However, a slightly lower slope of $v_r - \Delta P$ curves for 100S and 200S PDMS microchannels can be attributed to the slightly larger (<10%) hydraulic diameters of PDMS channels compared to the glass channels (Table S1). A slightly higher slope of the $v_r - \Delta P$ curve for the 500S PDMS channel compared to the 500S glass channel is due to a comparable contribution of the softer PDMS microchannel compliances with respect to the syringe holder compliance to the total fluidic capacitance of the circuit (Fig. S4). On the other hand, for the SC-HSF configuration, the softer PDMS channels had a higher contribution to the fluidic circuit capacitance compared to that of the connecting tubing, therefore, the $v_r - \Delta P$ curves for the PDMS S100 and S200 microchannels were much higher than those for the glass microchannels of similar diameter (Fig. 4b and **Fig. S5**). Shaded areas show the difference in $v_r$ values for the PDMS and glass microchannels at any given $\Delta P$. The response of 500S PDMS and glass microchannels is very similar despite the deformable walls of the former channel. The 500S PDMS channel has a much higher volumetric capacity and fluidic capacitance compared to the 100S and 200S PDMS channels, where relatively higher $v_r$ values could be expected compared to a glass channel (Fig. 4c and Fig. S5). However, the flow resistance of the 500S PDMS microchannel is much lower than the 100S and 200S microchannel, therefore, the stored fluidic capacitance within the PDMS channel (*viz.* absent in the glass channel) was quickly dispersed throughout the circuit as the pressure was homogenized from inlet to outlet. Since the residual velocity $v_r$ was measured only after 2 s of initiating the flow stoppage, much of the fluid flow associated with the high compliance of the 500S PDMS channel was not accounted for, therefore, a relatively low residual velocity was recorded (Fig. 4b and Fig. S1b). As the SC-HSF is operated at high pressure, the PDMS walls stay deformed, even after $\Delta P$ is removed, to retain some of the fluid volume inside the channel that would have traveled from the inlet tubing toward the outlet tubing during the stop flow (Fig. 4c and **Table S4**).

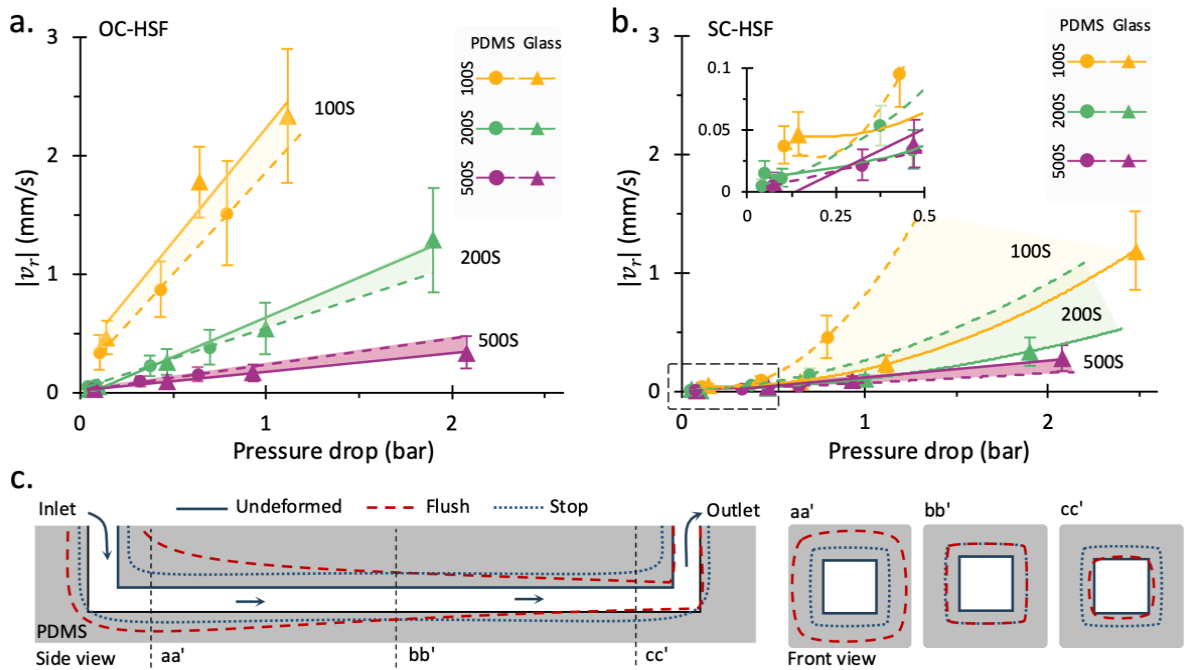

**Figure 4. Comparison of PDMS and glass microchannels operated with OC-HSF and SC-HSF configurations.** (a) The PDMS and glass microchannels response is very similar under OC-HSF conditions for all the diameters. (b) The 100S and 200S PDMS channels used with SC-HSF configuration resulted in much higher $v_r$ values compared to the glass channel. The response for 500S PDMS and glass channels was very similar. Shaded regions between lines indicated the difference

between the soft and hard wall channels (PDMS and glass) for each channel size using OC-HSF (a) and SC-HSF (b) configurations. Data points represent means of multiple cycles and particle replicates ($N_{cycle}$ = 3, $N_{particle}$ > 100) and error bars show standard deviations. Lines represent linear fits to the OC-HSF data (a) and 2nd order polynomial fits to the SC-HSF data (b). (c) Schematic of a PDMS microchannel shows cross-sectional side and front views to highlight the deformation of the channel walls during the flush and stop phases of the stop-flow cycle.

**Numerical model estimates the fluidic capacitance in the circuit and validates the experimental results by matching the sample's traveling distances:**

A numerical model couples the fluid dynamics and solid mechanics physics to simulate the expansion of elastic tubing and microchannel connected in the fluidic circuit for a given inlet pressure (**Fig. 5**). The pressure gradient results in a maximum expansion of the inlet tubing and a capacitive volume produced upstream of the microchannel that will flow through the microchannel to the downstream outlet tubing to homogenize the pressure difference (Fig. 5a). For a comparison between the experimental and numerical results for the SC-HSF configuration, the average velocity curve is integrated over time (2-47 s) to measure the total distance traveled ($T_{D,Exp}$) by the media as the flow is fully stopped (Fig. S1b). A correction factor ($\chi$) is numerically evaluated for each channel size to account for the overestimation of average residual velocity in the microchannel due to the limited depth of field of the objective (Fig. S1c). For the numerical estimation of the travel distance ($T_{D,Num}$) lapsed by the medium, the overall volume expansion ($\Delta V_{Num}$) is calculated by using the Fluid-Structure-Interaction (FSI) simulation model. We assumed that half of this expanded circuit volume would have to pass through the microchannel of a given cross-section area ($S$) to balance the pressure gradient, i.e., 0.5 x $\Delta V_{Num}$ = $S$ x $T_{D,Num}$. Expanded cross sections of the inlet tubing (aa'), microchannel (bb', and cc'), and outlet tubing (dd') indicate the change in diameters ($\Delta D$) of a respective cross-section under a given pressure, which is used to calculate $\Delta V_{Num} \sim \Delta D_{Num}^2$ x $L_C$, where $L_C$ is tubing or channel length (Fig. 5a). Here the numerical model directly provides the $\Delta V_{Num}$ values. A reasonable agreement between the $T_{D,Num}$ (= 0.5 x $\Delta V_{Num}$ / $S$) and $T_{D,Exp}$ (= $\chi \int v_r \, dt$) is found for both the glass and PDMS microchannels (Fig. 5b). In the SC-HSF configuration using glass microchannels, the tubing expansion is primarily contributing towards capacitive residual flow. For the 100S and 200S PDMS channels, tubing contribution is still dominant; however, the expansion of the 500S microchannel starts to contribute equally as the channel diameter approaches the tubing diameter (Fig. S5).

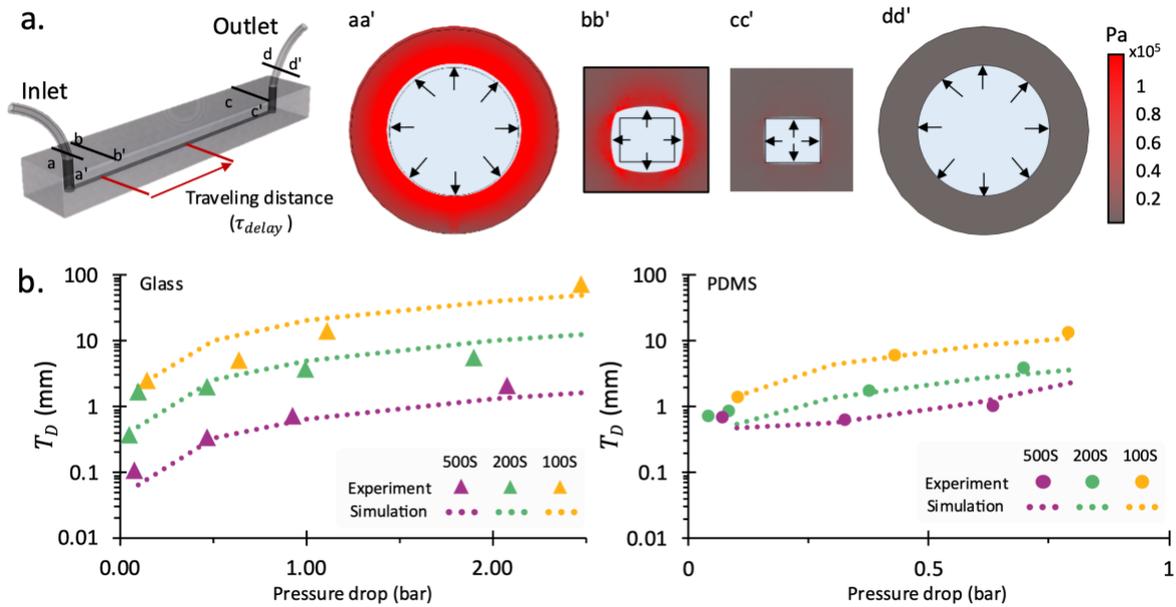

**Figure 5. Simulated and measured traveling distances during stoppage sequence.** (a) 3D numerical simulation performed on tubing and soft wall PDMS microchannel (at left); aa', bb', and cc' show expansion under applied fluid pressure for tubing, the channel', respectively. (b) Traveling distances after stopping fluid flow in soft PDMS (right) and hard glass (left) channels for various cross-sectional dimensions. Data points represent indirectly calculated displacements of particles from measured residual velocity in the SC-HSF configuration. Dotted lines represent the traveling distances calculated by using the numerical model.

## Conclusions

A steady microfluidic flow through a microchannel can be characterized by the following four parameters: (i) channel dimensions, (ii) sample viscosity, (iii) flow rate, and (iv) pressure gradient across the channel length, which were correlated by the Hagen-Poiseuille equation. However, during a transient phase of stopping the microfluidic flow, an additional parameter in the form of (v) microchannel compliance or elasticity was considered. The flow inside the microchannel was stopped as the pressure gradient was neutralized by one of the three stop flow configurations: LSF, OC-HSF, and SC-HSF (Fig. 1). The stopped-flow was characterized by a residual velocity inside the microchannel while the pressure gradient was homogenized. The residual velocity was an outcome of the inhomogeneous compliance of the fluidic circuit, i.e, the microchannel walls and the connecting tubings, due to the pressure gradient across the fluidic circuit. The pressure gradient was linearly correlated with the flow rate, where the proportionality constant in the form of hydraulic flow resistance included the sample viscosity and the microchannel dimensions. Therefore, the residual velocity of a stopping flow was investigated for the variable (1) pressure gradients, (2) channel diameters, (3) channel and tubing materials (fluidic circuit compliance), and (4) sample viscosities.

1. The residual velocity increased linearly and quadratically with the pressure gradient (0 - 2.5 bar) for the OC-HSF and OC-HSF configurations, respectively (Fig. 2b-c). The residual velocity in the LSF configuration was independent of pressure drop for smaller microchannels; however, the residual velocity increased with pressure drop for the larger microchannel as the contribution of the hydrostatic pressure in the system became significant (Fig. 2a,d,e).
2. The residual velocity decreased with the microchannel diameter (100 - 500 μm) following an inverse power law with exponents ~-1.7 and ~-0.4 for the OC-HSF and SC-HSF configurations, respectively (Fig. 2f). For the LSF configuration, the residual velocity increased with the microchannel diameter following a power law with exponent ~3.

It was observed that the LSF configuration was strongly influenced by the hydrostatic pressure fluctuations within the larger diameter channels; therefore, the LSF configuration was not considered for additional parametric investigations.

3. The residual velocity was measured within PDMS and glass microchannels by using the OC-HSF and SC-HSF configurations (Fig. 4). For the SC-HSF configuration, a higher compliance of the PDMS channels resulted in relatively higher residual velocities compared to the glass channels. For the OC-HSF configuration, the difference in the residual velocities for different channel materials was not significant as the compliance of the syringe holder overshadowed the microchannel compliance. Overall, the SC-HSF resulted in lower residual velocities compared to OC-HSF (e.g. 12x lower for 100S channel at 1 bar, (Fig. 1f)). Moreover, the SC-HSF configuration was tested with different connecting tubing materials of variable elastic moduli (10 MPa to 180 GPa), where the residual velocity was higher in the softer PVC tubing compared to harder steel tubing, i.e. ~5x higher at 1 bar.
4. The residual velocity was also measured for three different sample viscosities (1 – 24 mPa.s) by the SC-HSF configuration (Fig. 3a). The residual velocity decreased with the sample viscosity following an inverse power law with exponent -1.29, i.e. 53x reduction in residual velocity as the viscosity was increased from 1 to 24 mPa.s (Fig. S4). This indicated that the high-viscosity sample damped the residual flow faster.

Finally, a numerical simulation model was developed to understand the contributions of individual elements of the fluidic circuit to the residual flow under the SC-HSF configuration (Fig. 5). The numerical model simulated the expansion of the tubing and microchannel due to the pressure gradient, and provided the capacitive volume of the fluidic circuit. A traveling distance was calculated based on the experimental residual velocities and the numerically estimated capacitive volumes. The numerical and experimental traveling distances matched well, thereby, confirming our hypothesis that the residual flow within the microchannel was primarily due to the compliance of the fluidic circuit.

In summary, the LSF configuration performed best with the high flow resistance microchannels (~$10^{14}$ Pa.s/m$^3$) but the performance decreased with the flow resistance. The OC-HSF and SC-HSF configurations resulted in the least residual velocities inside the low flow resistance microchannel (~$10^{11}$ Pa.s/m$^3$), where the performance decreased with the flow resistance. The SC-HSF configuration performed better than the OC-HSF configuration in general.

# Supporting Information

**Materials and Methods:**

**Microfluidic devices:** Glass capillaries (VitroTubes™, 8250, 8320, and 8510, 50mm in length, VitroCom) were sealed at both sides with 18G blunt-end needles by using two-component epoxy glue (UHU[R], 2-component epoxy adhesive, Bolton Adhesive) to realize glass microchannels. The PDMS microchannels were fabricated by using the soft lithography process. The master molds for the 100S and 200S microchannels were prepared by spin coating a photoresist (SU8 2100, Microresist Technology Gmbh) on a silicon wafer (Microchemicals Gmbh) and UV patterning it by a laser writer (Dilase 250, KLOE Design). The master mold for the 500S PDMS microchannel was printed by a stereolithography (SLA)-based 3D printer (Sonic mini 8K, Phrozen Technology). The PDMS microchannels' dimensions after the replica molding process are provided in the supporting information (Table S1).

**Experimental setup:** The microfluidic devices (i.e., PDMS or glass microchannels) were connected to the syringe pump (Nemesys S, CETONI Gmbh) or the pneumatic pump (Flow EZ™ 7000, FLUIGENT Gmbh) at the inlet by using the PTFE tubings (1/16" – 1 mm OD-ID, TechLab), unless otherwise mentioned. A reservoir tube fitted with a gas-tight p-cap (P-cap 15 mL, FLUIGENT Gmbh) was required for the pneumatic pump. A collection tube (15 mL PP tube, CELLSTAR®, Greiner Bio-One International GmbH) was connected at the outlet of the microchannel through a valve (ASCO 833-630887 3-way vale 3 bar, EMERSON Electric Co.). For experiments with different inlet tubings, PVC-based (Tygon® ND 100-80 Tubing, 1.78 - 1.02 mm OD-ID, Saint-Gobain Performance Plastics Corporation) and steel-based (Steel capillary, 1/16" – 1 mm OD-ID, CETONI Gmbh) tubings with the similar dimensions were used instead of PTFE tubing.

**Particle velocity measurement:** The central region of the microchannels (region of intertest (ROI): channel width x 2500 µm length) was observed by a microscope (Thunder Imager DMi8, Leica Microsystems) with an integrated camera (DFC9000, Leica Microsystems) (Fig. S1a). An image was captured every 40 ms (i.e., 25 fps) to record the motion of 10 µm polystyrene green fluorescent microparticles. After waiting for at least four stop-flow cycles to pressurize the system, the particles were subsequently recorded for three consecutive cycles. The particles were segmented and their positions were determined for each frame to calculate their velocities by using an in-house built MATLAB code based on Kalman filtering. After determining the individual particle velocities for each cycle, the velocity data was overlaid for the three cycles to take an average residual velocity (Fig. S1b).

**Particle velocity correction based on the depth of field:**

The particles flowing through the microchannels (100S, 200S and 500S) were imaged within the ROI by using a 5x objective, whose depth of focus is defined as[26]:

$$d_{dof} = (\lambda \cdot n) \div NA^2 + (n \cdot e) \div (M \cdot NA)$$

where $\lambda$ light wavelength (470nm), $n$ is the refractive index of the medium (1.0 for air, dry objective), $NA$ is the numerical aperture (0.12), $M$ is the objective magnification (5) and $e$ is the smallest distance that can be resolved by a detector (6.5 µm). Since the calculated $d_{dof}$ (~45 µm) was smaller than the microchannel heights (100 µm or greater), a portion of the randomly distributed particles within the microchannel cross-section was not perfectly focused within the acquired images. For the 100S microchannel, a small portion of the slightly defocused particles within the top and bottom ~27 µm height of the microchannel was accounted for in the velocity calculations by adjusting the threshold parameters. However, for the 200S and 500S microchannels, a significantly large portion of the particles was not captured with sufficient fluorescent intensity withing the defocused microchannel heights of ~100 µm and ~400 µm, respectively. Therefore, it was essential to define a correction factor ($\chi$) that accounted for the defocused particles to calculate the average velocities within the larger 200S and 500S microchannels. The correction factor was calculated from the simulation results as follows: $\chi = v_{avg}/v_{avg,DOF}$ (Fig. S1c). For the experimental results, a threshold value (smallest possible) was set based on the microparticle images acquired within the 100S microchannel for creating a mask to segment all the particles across the 100 µm height. For the similar excitation intensity and microscope parameters, this threshold value was applied to the images acquired for the 200S and 500S microchannels to obtain the particle velocities within the middle 100 µm height of these channels (Fig. S1c). The corrected average velocities were obtained by multiplying the above measured particle velocity values within 200S and 500S channels with the correction factors $\chi_{200S}$ = 0.75 and $\chi_{500S}$ = 0.72, respectively.

**Inlet pressure measurement:** Pressure drop $\Delta P$ over the channel length was measured by an in-house built setup in which a pneumatic pump with a pressure controller (Flow EZ[TM] 7000, FLUIGENT Gmbh) was connected to the syringe pump by a 3-way connector (Fig. S2a). The inlet pressure at the syringe pump was measured by regulating the pneumatic pressure to keep the air-liquid interface in the branch tubing preserved at a certain flow rate.

**Viscosity measurement:** Hagen-Pouseille equation, which relates the pressure drop with flow rate for incompressible Newtonian fluids in a laminar flow regime, was used to indirectly measure the sample viscosities:

$$\Delta P = \frac{8\mu L Q}{\pi R^4}$$

where $\Delta P$ is pressure drop, $Q$ is the flow rate, $\mu$ is the viscosity of the fluidic medium, $L$ and $R$ are length and radius of the pipe, respectively. Here, we are taking a simplistic assumption that the sample fluid is Newtonian in nature, even though it has a variable viscosity with applied shear rate. A tubing, with known diameter and length, connected a pressurized reservoir to a collection tube placed on an analytical scale to measure the weight of the dispensed fluid over time (Fig. S2b). This provides an average flow rate ($Q$) through the tubing with a known pressure drop $\Delta P$ and flow resistance ($\frac{8\mu L}{\pi R^4}$). Therefore, the viscosity of the sample fluidic could be calculated by using the Hagen-Pouseille equation. The shear rate ($\gamma$) was calculated by: $\gamma = \frac{4Q}{\pi R^3}$. In-lab measurement of sample viscosities over variable shear rates resulted in a constant viscosity for water and decreasing viscosity values for the polymer solutions, which confirmed Newtonian and non-Newtonian shear-thinning behaviors, respectively (Fig. S2b). The shear-thinning nature of polymer solutions[27] was confirmed as the applied shear rate (50-7500 s$^{-1}$) resulted in ~30% and ~15% reduction in the viscosities for the 50% and 100% polymer solutions, respectively. Our viscosity measurements (Fig. S2b) were similar to those reported earlier[28]; however, they differed from the manufacturer (Sigma-Aldrich)'s datasheet (57 mPa for the pure PEGDA (M$_n$=575)). This difference in measured viscosities could be due to the variable room temperatures or the applied shear rates as also indicated by others.[29]

**Residual pressure measurement:** It is generally understood that the pressure inside the fluidic circuit stays close to the inlet pressure as the flow is stopped and the pressure gradient is removed by using the HSF configurations.[22], [23], [30] However, this has not been experimentally verified that how different this residual pressure during the stop phase is with respect to the maximum inlet pressure during the flush phase. An experimental setup was built to measure this residual pressure ($P$) during the SC-HSF configuration by tracing the length $L$ of a trapped bubble to estimate its volume ($V$) inside of a side tubing (Fig. S2c). By using the ideal gas equation, $PV = nRT$, the volume change of the trapped bubble was related to the pressure change inside the fluidic circuit (Table S5). Under the same environmental conditions, i.e., constant room temperature $T$, the equation can be written for the three states of the bubble as: $P_1 V_1 = P_2 V_2 = P_3 V_3 \rightarrow P_1(\pi R^2 L_1) = P_2(\pi R^2 L_2) = P_3(\pi R^2 L_3)$. The radius $R$ of the tubing can be considered constant, therefore, the above equation can be written as: $P_1L_1 = P_2L_2 = P_3L_3$. The three states of the bubble are described as: (1) the initial state before the start of the stop flow cycle, (2) the high pressure state during the flush phase of the cycle, and (3) the residual pressure state during the stop phase of the cycle. Table S5 summarizes the measured lengths of the bubble under three states and the associated bubble pressure calculated from the above equation. It is concluded that the residual pressure in the fluidic circuit was nearly half the pressure during the flush phase of the stop-flow cycle. This confirms the graphical representation of the stop flow cycle for the HSF configurations in Fig. 1.

**A numerical model for the expansion of elastic components:** Fluid-structure interaction model was used to couple the 'solid mechanics' and 'laminar flow' physics (COMSOL Multiphysics) to determine the expansion of elastic components in the fluid circuit due to a pressure gradient. For the expansion of inlet tubing (PTFE, density: 2200 kg/m$^3$, Young modulus: 0.4 GPa, Poisson ratio: 0.4), a single phase laminar flow (fluid density: 1060 kg/m$^3$, viscosity: 7.6 mPa.s) was simulated. The inlet boundary condition was a uniform flow rate of 0.001 µL/s, whereas the outlet boundary condition was set to be a static pressure of 0, 1, or 2 bar for a 10 mm long PTFE tubing section. The pressure drop $\Delta P$ across the PTFE tubing with 1 mm inner diameter was not significantly high compared to the $\Delta P$ across the 100S or 200S microchannels with relatively high flow resistances (Fig. S2c). Therefore, the expansion of the tubing across its length was relatively uniform. The capacitive volume for the entire inlet tubing length was calculated by linearly extrapolating the expansion of the 10 mm segment to a 40 cm long segment for glass channel and a 35 cm long segment for PDMS channel. However, when the tubing diameter (1 mm) was comparable with the inner dimension of the 500S microchannel, the $\Delta P$ across the inlet or outlet tubing length was no longer minuscule compared to the $\Delta P$ across the microchannel. Therefore, the expansion of tubing cannot be considered uniform across its length but in the form of a conical frustum. Simulating an entire 40 cm long tubing would be really challenging to directly obtain the expanded volume. Instead, the expansion at the starting and ending segments (10 mm long) of the tubing was obtained from the simulation, and the expanded volume was calculated for a frustum shape.

**Sample preparation:** Poly(ethylene glycol) diacrylate (PEGDA Mn:575, Sigma-Aldrich) was diluted with filtered DI water (50% volume ratio) for all the experiments except the investigation of sample viscosity effect on residual flow velocity. The solution was mixed with fluorescent tracing particles (PS-FluoGreen-10.0, Microparticles Gmbh) at ~10$^3$ particles per milliliter.

**Supporting figures:**

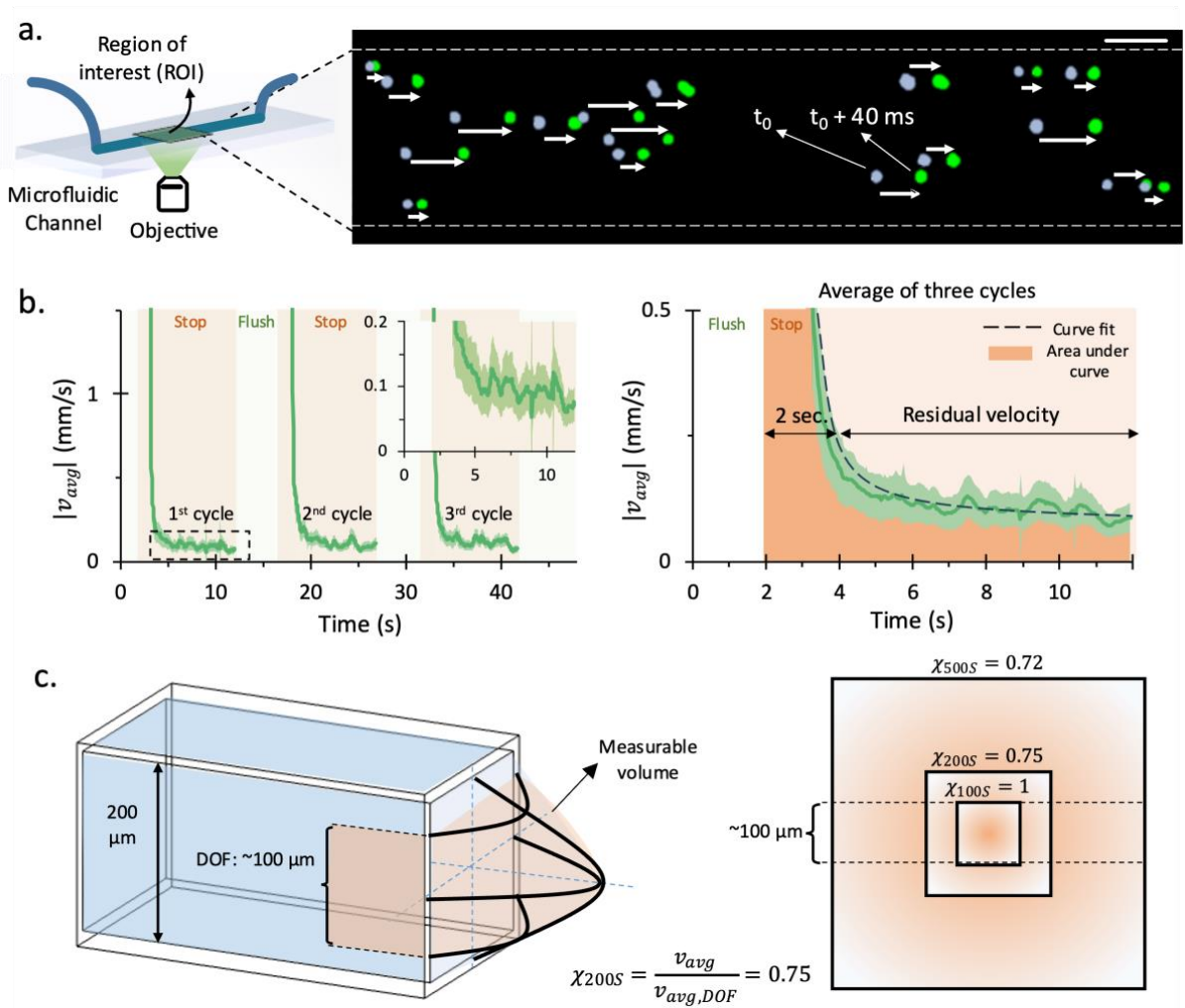

**Figure S1. Residual velocity measurement by particle tracing.** (a) Schematic of data collection unit (left) and an instance from the collected frames of the microparticles flowing through the microchannel (right). Overlapped image shows two sequential frames stacked together, where the blue and green colored particles indicate time steps $t_0$ and $t_0$ + 40ms, respectively. The white arrows indicate the velocity magnitude of each particle. The scale bar is 50 µm. (b) Particle velocities indicate the average residual flow velocity ($v_r$) during the stop phase (shaded region) of the three stop-flow cycles (left). Particle velocities averaged over the three cycles to determine the residual velocity within the microchannel over time (right). A second-order exponential fitted curve is integrated over time to obtain the distance traveled by the fluid during the stop phase of the cycle. (c) The depth of focus (DOF ≅ 100µm) of the objective determines the acquisition field through which the flowing particles are measured. A velocity correction factor ($\chi$) is calculated for all the microchannels by dividing the average velocities over the channel cross section ($v_{avg}$) and over the DOF rectangular area ($v_{avg,DOF}$). The average velocities are obtained from the numerical simulations.

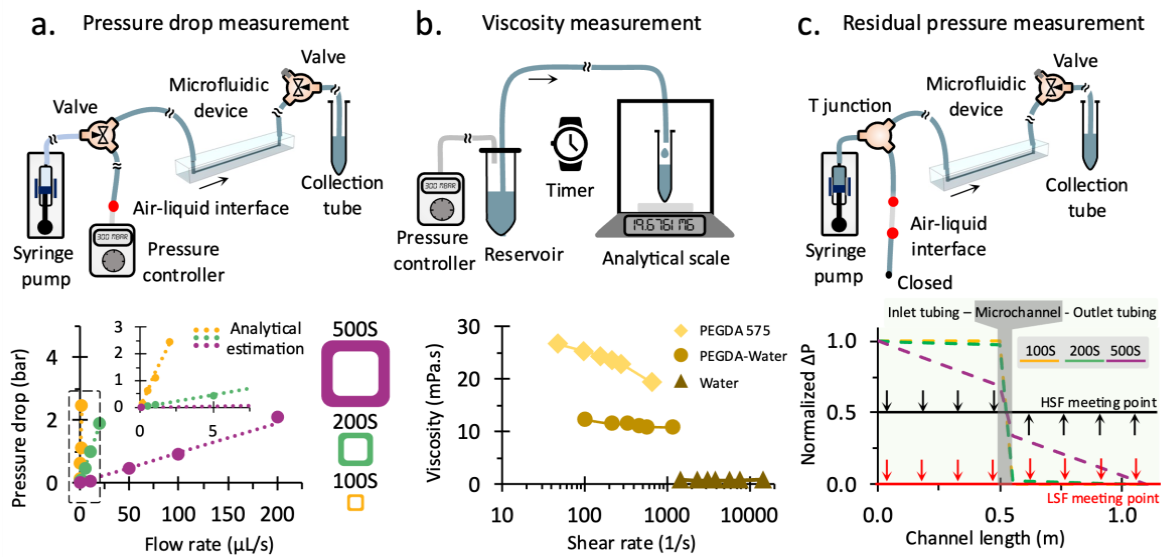

**Figure S2. Measurement of pressure drop, viscosity, and residual pressure.** (a) A schematic of the experimental setup to measure the inlet pressure at variable flow rates ($Q$) through different microchannels. The measured values (solid circles) of inlet pressure linearly correspond to the flow rates set at the syringe pump. The dotted lines indicate the analytically calculated pressure drop over the microchannel. The slope of the $\Delta P - Q$ lines decreases as the microchannel diameter is increased from 100S to 500S. (b) A schematic of the experimental setup to measure the viscosity of aqueous-polymer mixtures, i.e. water, 50% PEGDA solution in water, and 100% PEGDA. The viscosity is calculated by using the Hagen-Poiseuille equation for a given pressure drop and flowrate through tubing with known dimensions. The lowest measured viscosity values for water stay constant for the variable shear rates applied (i.e. increasing flow rate), confirming Newtonian fluid-like behavior of water. The viscosity values measured for the 50% and 100% PEGDA solutions decrease with increasing shear rate, which indicates the non-Newtonian fluid-like behavior of these polymer solutions. (c) A schematic of the experimental setup to measure the residual pressure within the fluidic circuit as the flow is stopped. A bubble trapped within a closed-end branch tubing results in two air-liquid interfaces, which are traced as the pressure within the main fluidic circuit fluctuates. The bottom interface does not move significantly as the liquid portion next to the closed end is nearly incompressible. The top interface moves down as the pressure within the fluidic circuit increases, thus compressing the trapped air bubble. The change in bubble volume is related to the pressure change in the circuit. (Bottom) Normalized pressure drop ($\Delta P_N$) across the fluidic circuit for different microchannels is plotted. The $\Delta P_N$ is minimal across the inlet and outlet tubings (1 mm in diameter) compared to that across the 100S and 200S microchannels. The $\Delta P_N$ across the 500S microchannel is of a similar order as that across the tubings. For the LSF configuration, the overall residual pressure drops to 0 (atmospheric pressure). For the HSF configurations, the residual pressure is averaged out to an intermediate value above the atmospheric pressure but lower than the highest pressure within the fluidic circuit during the flush phase of the cycle.

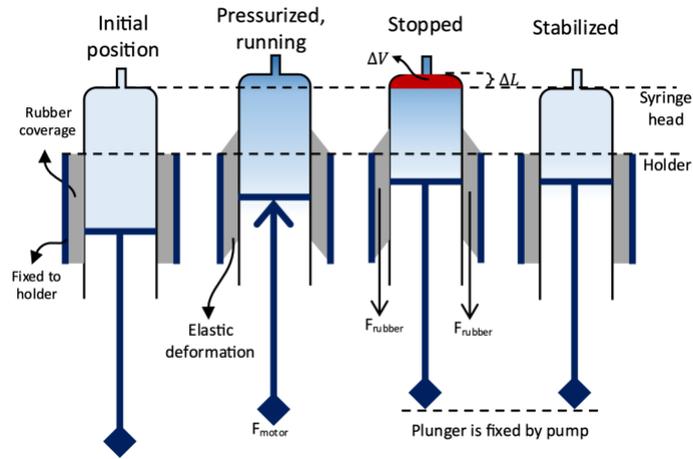

**Figure S3.** Compliance in the glass syringe holder padding contributes to the residual velocity within the microchannel for the OC-HSF configuration.

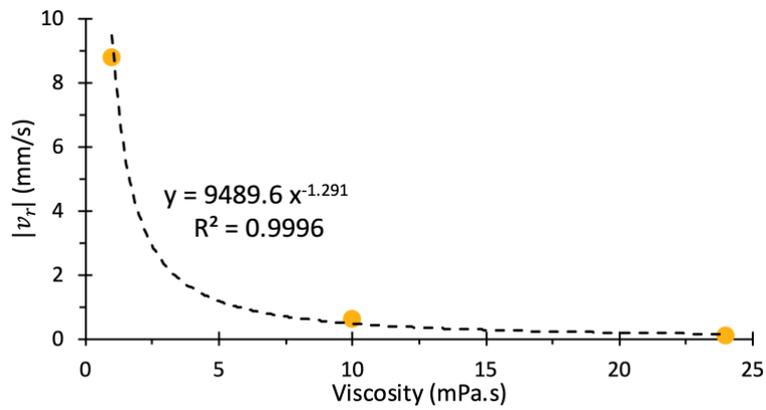

**Figure S4.** Residual velocities by varying medium viscosity within the microchannel for SC-HSF configuration

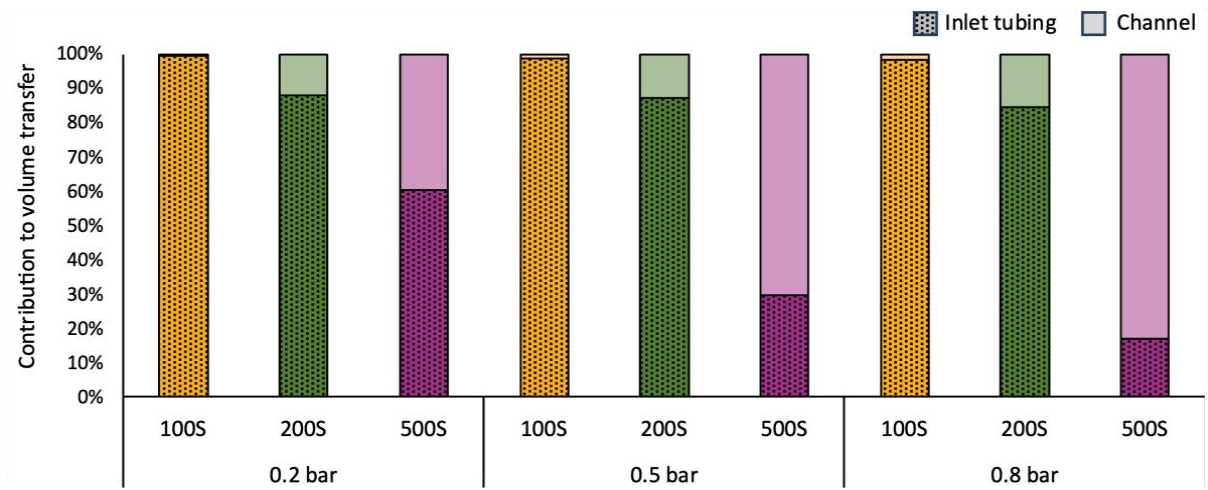

**Figure S5.** Contribution of PDMS channel and PTFE inlet tubing with respect to different pressure and crossectional areas.

**Supporting tables:**

**Table S1.** Microchannel dimensions and material properties

| | Microchanel material | | | | | |
|---|---|---|---|---|---|---|
| | Glass | | | PDMS | | |
| Dimensions | 100S | 200S | 500S | 100S | 200S | 500S |
| **Width (µm)** | 100∓10 | 200∓20 | 500∓50 | 119.5∓0.34 | 212∓1.41 | 611∓11.03 |
| **Height (µm)** | 100∓10 | 200∓20 | 500∓50 | 97.15∓1.37 | 195.4∓1.13 | 490∓0.28 |
| **Length (cm)** | 5∓0.2 | 5∓0.2 | 5∓0.2 | 5∓0.1 | 5∓0.1 | 5∓0.1 |
| **Young Modulus (GPa)** | 73.1 | | | 0.00075 | | |
| **Poisson ratio** | 0.17 | | | 0.49 | | |

**Table S2.** Tubing properties

| Material | PVC | PTFE | Steel |
|---|---|---|---|
| **Inner diameter (mm)** | 1∓0.02 | 1∓0.02 | 1∓0.02 |
| **Outer diameter (mm)** | 1.6∓0.02 | 1.6∓0.02 | 1.6∓0.02 |
| **Inlet length (cm)** | 35∓0.5 | 35∓0.5 | 35∓0.5 |
| **Outlet length (cm)** | 40∓0.5 | 40∓0.5 | 40∓0.5 |
| **Young Modulus (GPa)** | 0.01 | 0.4 | 180 |
| **Poisson ratio** | 0.38 | 0.46 | 0.3 |

**Table S3.** Sample properties

| | PEGDA | PEGDA-Water (1:1) | | Water | |
|---|---|---|---|---|---|
| **Density (kg/m$^3$)** | 1120 | 1060 | | 1000 | |
| **Average Viscosity (mPa.s)** | 26.6 | 10 | | 1 | |
| **Suspended particle** | Material | Polystyrene | Density (kg/m$^3$) | 1050 | Diameter (µm) | 10 |

**Table S4:** Measured lengths of trapped bubble and absolute (relative) pressure values of different states of flow condition following: $P_1 L_1 = P_2 L_2 = P_3 L_3$. $P_m$ is the measured pressure at a given flow rate $Q$ (Fig. S2a).

| $Q$ (µL/s) | $P_m$ (bar) | L1 (mm) | P1 (bar) | L2 (mm) | P2 (bar) | L3 (mm) | P3 (bar) | Ratio (P3/P2 & L2/L3) |
|---|---|---|---|---|---|---|---|---|
| 5 | 1.48 (0.48) | 70.0 | 1 (0) | 47.1 | 1.45 (0.45) | 53.0 | 1.30 (0.30) | 0.66 |
| 10 | 1.94 (0.94) | 70.0 | 1 (0) | 33.8 | 2.07 (1.07) | 45.9 | 1.53 (0.53) | 0.49 |

Time scales associated with flow stoppage and capacitance in the fluidic circuit:

Following Hagen–Poiseuille law, an incompressible, Newtonian, viscous fluid flowing through a confined microchannel due to a pressure gradient ($\Delta P = P_{in} - P_{out}$) present across the flow path, should ideally stop immediately as soon as $\Delta P \to 0$. However, in reality, the flow only stops after a certain delay time, which could possibly be attributed to (1) a pressure wave traveling across the microchannel, (2) an inertia of the moving fluid, or (3) a compliance within the fluidic circuit.[31]

(1) Assume a water-filled microfluidic channel (15 cm long) with a certain $\Delta P$ across its ends. As soon as the pressure source is removed from the inlet, a pressure fluctuation in the form of a sound wave could propagate across the length of the microfluidic channel within a time period of $\tau_p = 0.15\, m\, /\, 1500\, ms^{-1} = 100$ µs. This indicates that the pressure pulse could bounce off the ends of the channel for 10,000 times within a second, and will be totally lost to the viscous damping within the fluid during this time. Therefore, we deduce that the effect of such pressure fluctuations would be short-lived and would not contribute significantly to the bulk fluid movement as in the case of an acoustic streaming flow.[11]

(2) A non-zero residual velocity during the stop-flow could be associated with the inertia of the fluid until internal viscous losses diminish the kinetic energy of the flow. The viscous damping of the inertial forces is described by an inertial time constant $\tau_i = \frac{1}{2.405^2} T_0$, where $T_0 = a^2/\upsilon$ is the momentum diffusion time, $a$ is the diameter of the circular channel, and $\upsilon$ is kinematic viscosity.[31] Drawing from an analogy of an electrical capacitor charging/discharging, a 99.99% damping of fluidic inertia would reach within 10 x $\tau_i$. For water ($\upsilon_{water} \simeq 1\, mm^2/s$) flowing inside a 200 µm channel (15 cm long), the time constant $\tau_i$ is calculated as 7 ms, where a flow will lose 99.99% of its kinetic energy within 70 ms. That means, a 5 mm/s ($Re = 1$) flow of water will reach a residual velocity of 0.5 µm/s in 70 ms, whereas a 500 mm/s ($Re = 100$) flow will have a residual velocity of 50 µm/s after the same delay time. However, we have observed experimentally that the water flowing through a 200S microchannel has a residual velocity of the order of mm/s even after a delay time of the order of seconds (Fig. 3a). Therefore, we can deduce that the inertial effect diffuses too fast to contribute significantly to the residual velocity of an inertial flow.

(3) A fluidic circuit with non-rigid walls will have a fluidic capacitance to store potential energy in the system as the elastic walls are stretched under high pressure. The stored potential energy then drains away in the form of the kinetic energy of the residual flow as the elastic walls relax back to the original state and pressure is homogenized across the fluidic circuit. A microchannel or tubing (diameter $D$, length $L$) with an elastic modulus $E$, deforming under an applied pressure, will have a relaxation time constant associated with fluidic circuit capacitance as $\tau_c \sim \mu L^2/ED^2$ for a fluid with viscosity $\mu$. The relaxation time $\tau_c$ can vary from as low as 100 ms to as high as 100 s depending upon the microchannel material.[12] For example, for a PDMS microchannel (PVC tubing; $E$ ~10MPa) will have a 3-4 orders of magnitude higher $\tau_c$ compared to a glass microchannel (steel tubing; $E$ ~70GPa) with similar dimensions, which is a consistent with the experimental observations (Fig. 3b and Fig. 4c). However, despite of multiple orders of magnitude difference in $\tau_c$ values, the relative residual velocities in a PDMS/glass microchannel (or with PVC/steel tubings) differed by only a multiple factors of ≲8x. This inconsistency can be explained by the fact that the residual velocities are only measured after a certain delay time of 2 s, where much of the relaxation of microchannel or tubing has already happened (Fig. S1b).